\documentclass[twocolumn,preprintnumbers,superscriptaddress,amsmath,amssymb,prb]{revtex4-1}
\usepackage{braket}
\usepackage{miller}
\usepackage{changes}

\usepackage[textsize=tiny]{todonotes}
\usepackage{graphicx}

\usepackage[utf8]{inputenc}
\usepackage{multirow}
\usepackage{float}
\usepackage{bm}
\usepackage{verbatim}
\usepackage{xfrac}
\usepackage{array}
\usepackage{siunitx}
\usepackage{ulem}

\begin{document}
	
	\title{Orbital angular momentum bistability in a microlaser}
	
	\author{N. Carlon Zambon}
	\thanks{These authors contributed equally to this work.}
	\affiliation{Centre de Nanosciences et de Nanotechnologies (C2N), CNRS - Université Paris-Sud / Paris-Saclay, Marcoussis, France}
	
	\author{P. St-Jean}
	\thanks{These authors contributed equally to this work.}
	\affiliation{Centre de Nanosciences et de Nanotechnologies (C2N), CNRS - Université Paris-Sud / Paris-Saclay, Marcoussis, France}
	
	\author{A. Lemaître}
	\affiliation{Centre de Nanosciences et de Nanotechnologies (C2N), CNRS - Université Paris-Sud / Paris-Saclay, Marcoussis, France}
	
	\author{A. Harouri}
	\affiliation{Centre de Nanosciences et de Nanotechnologies (C2N), CNRS - Université Paris-Sud / Paris-Saclay, Marcoussis, France}
	
	\author{L. Le Gratiet}
	\affiliation{Centre de Nanosciences et de Nanotechnologies (C2N), CNRS - Université Paris-Sud / Paris-Saclay, Marcoussis, France}
	
	\author{I. Sagnes}
	\affiliation{Centre de Nanosciences et de Nanotechnologies (C2N), CNRS - Université Paris-Sud / Paris-Saclay, Marcoussis, France}
	
	\author{S. Ravets}
	\affiliation{Centre de Nanosciences et de Nanotechnologies (C2N), CNRS - Université Paris-Sud / Paris-Saclay, Marcoussis, France}
	
	\author{A. Amo}
	\affiliation{Physique des Lasers, Atomes et Molécules (PhLAM), CNRS - Université de Lille, Lille, France}
	
	\author{J. Bloch}
	\affiliation{Centre de Nanosciences et de Nanotechnologies (C2N), CNRS - Université Paris-Sud / Paris-Saclay, Marcoussis, France}
	
	\begin{abstract}
		Light's orbital angular momentum (OAM) is an unbounded degree of freedom emerging in helical beams that appears very advantageous technologically. Using a chiral microlaser, i.e. an integrated device that allows generating an emission carrying a net OAM, we demonstrate a regime of bistability involving two modes presenting distinct OAM ($\ell=0$ and $\ell=2$). Furthermore, thanks to an engineered spin-orbit coupling of light in the device, these modes also exhibit distinct polarization patterns, i.e. cirular and azimuthal polarizations. Using a dynamical model of rate euqations, we show that this bistability arises from polarization-dependent saturation of the gain medium. Such a bistable regime appears very promising for implementing ultrafast optical switches based on the OAM of light. As well, it paves the way to the exploration of dynamical processes involving phase and polarization vortices.
	\end{abstract}
		
		\maketitle
		
		\section{Introduction}
		Electromagnetic waves carry angular momentum through two main contributions: spin angular momentum (SAM) associated to circular polarization, and orbital angular momentum (OAM) emerging in beams presenting a helical phase front\cite{Allen1992}. While polarization is restricted to values of $\pm\hbar$, OAM is theoretically unbounded as it can take any value $\ell\hbar$, where $\ell$ is an integer corresponding to the number of times the phase front winds around the propagation axis within an optical period.
		
		Such an unbounded degree of freedom of light appears very advantageous technologically. Indeed, transferring arbitrarily large values of angular momentum to massive objectives is a powerful asset in opto-mechanics\cite{Aspelmeyer2014} and for optical trapping schemes\cite{Padgett2011}. Moreover, it could allow multiplexing classical\cite{Wang2012} or quantum information\cite{Molina-Terriza2007, Erhard2018} in higher-dimensional bases, thus enhancing the density and robustness of transmission channels.
		
		Fully taking profit of such high-dimensional bases requires the ability to manipulate OAM-carrying beam not only with linear optical elements, but as well in the nonlinear regime. Most notable demonstrations of nonlinear optical control of the OAM include the generation of higher-harmonics in nonlinear crystals\cite{Dholakia1996} and atomic vapors\cite{Kong2017}, and of OAM-entangled photon pairs by parametric down-conversion\cite{Mair2001}. Furthermore, recent demonstration of microlasers where the emission carries OAM with a chirality that can be optically controlled from clockwise to counter-clockwise vortices\cite{Zambon2018} offers new opportunities for exploring OAM-based nonlinear optics in integrated devices. 
		
		In this letter, we experimentally show that nonlinear effects associated to gain saturation in such microlasers lead to an optical bistability between modes presenting distinct values of OAM (i.e., $\ell=0$ and $+2$). Moreover, the engineered spin-orbit coupling of light in these devices allows switching not only the OAM magnitude of the beam, but also its polarization texture, from circularly to azimuthally polarized. This confluence of optical bistability and spin-orbit coupling of light is particularly interesting as it opens the door to the exploration of dynamical processes (e.g. quenches and phase transitions) involving distinct phase and polarization vortices\cite{Desyatnikov2005}.

		\begin{figure}
			\includegraphics[trim=0cm 0cm 0cm 0cm, width=85mm]{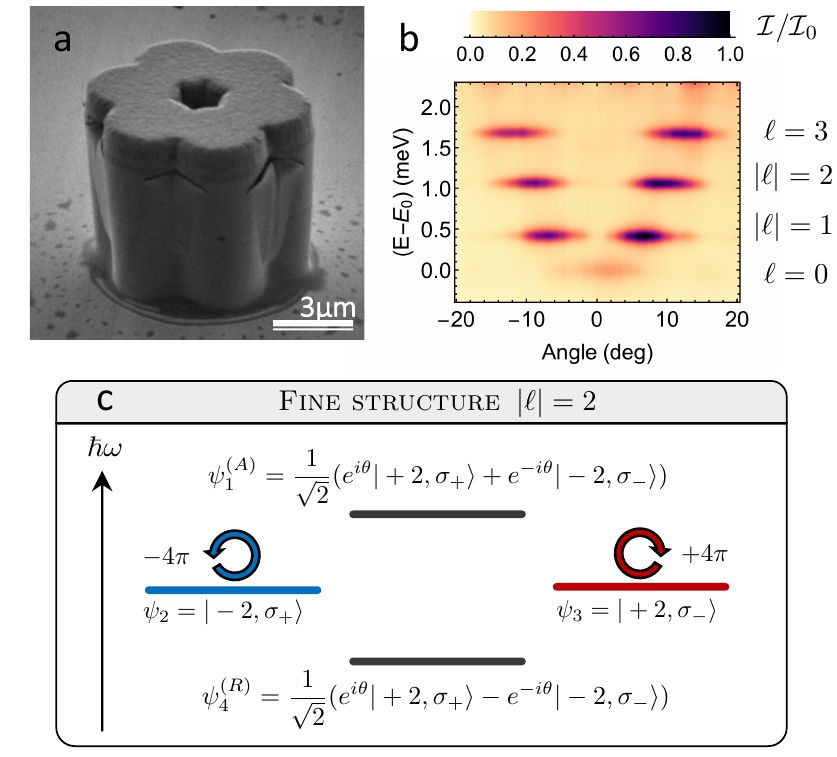}
			\caption{(a) SEM image of a device. (b) Angle-resolved emission spectrum of the molecule exhibits four energy levels corresponding to angular momenta $\ell=0,\pm1, \pm2 \;\mathrm{and}\; 3$. (c) Fine structure of the $|\ell|=2$ manifold in the presence of spin-orbit coupling. The middle states present a net OAM $\ell=\pm2$ and opposite polarizations: $\sigma_{+}$ (blue) and $\sigma_{-}$ (red).}
			\label{IntroFig}
		\end{figure}
		
		\section{Description of the chiral microlaser} 
		
		The chiral microlasers used in this work are built from semiconductor microcavities grown by molecular beam epitaxy. The cavities consist of a GaAs layer embedding a single $\mathrm{In_{0.04}Ga_{0.96}As}$ quantum well and inserted between two $\mathrm{Al_{0.95}Ga_{0.05}As/Al_{0.10}Ga_{0.90}As}$ Bragg mirrors formed from 36 (40) pairs in the top (bottom). To obtain microlasers with the appropriate discrete rotational symmetry for generating OAM, the cavities are processed by electron beam lithography and dry etching techniques to form hexagonal rings of coupled micropillars. Fig. \ref{IntroFig} (a) shows an electron microscopy image of the specific device used in this work (the pillars diameter is $\mathrm{3.2~\mu m}$ and the inter-pillar distance is $\mathrm{2.4~\mu m}$).
		
		Due to the discrete rotational symmetry of the microstructure, the photonic eigenmodes can be classified by their angular momentum $\ell$, which is associated to the evolution of the phase around the device\cite{Sala2015, Zambon2018}. In the tight-binding limit, this leads to the following four energy levels characterized by the quantum numbers $\ell=0,\pm1,\pm2,3$: 
		
		\begin{equation}
		\label{mode}
		\ket{\ell}=\frac{1}{\sqrt{6}}\sum\limits_{j}e^{2\pi i\ell j/6}\ket{\phi_{j}},
		\end{equation}
		
		\noindent where $\ket{\phi_{j}}$ corresponds to the ground state of the $j^{th}$ pillar. 
		
		States $\ket{\ell=0}$ and $\ket{\ell=3}$ do not carry angular momentum as their wave-function evolves respectively in- and out-of-phase between neighbouring pillars. On the other hand, states $\ket{\ell=\pm1}$ and $\ket{\ell=\pm2}$ carry a net angular momentum, corresponding to phase vortices of $\pm 2\pi$ and $\pm 4\pi$. These four energy levels are observed, well below the lasing threshold, with angle and energy resolved photoluminescence measurements (Fig. \ref{IntroFig} (b)).
		
		In order to generate a chiral emission, we take profit of a coupling between the spin and orbital angular momenta of photons that emerges in dielectric microcavities \cite{Sala2015,Dufferwiel2015,Bliokh2015}. This spin-orbit effect arises from an anisotropic inter-pillar coupling: the coupling energy is greater for photons polarized parallel to the axis linking two neighbouring pillars than for photons polarized perpendicularly\cite{MichaelisdeVasconcellos2011}. As a result of this azimuthally varying birefingent axis, the degeneracy of $\ell=\pm1$ and $\ell=\pm2$ manifolds is lifted resulting in a 3-level fine structures ($\ell=0$ and $\ell=3$ manifolds are not affected by this spin-orbit effect, as they do not carry OAM). These fine structures cannot be spectrally resolved below the lasing threshold (Fig. \ref{IntroFig} (b)), because the linewidth is larger than the energy spacing (related to the hopping anisotropy of $\sim20~\mu eV$); it can however be accessed in the lasing regime where the emission lines narrow significantly \cite{Sala2015, Zambon2018}.
		
		For the particular case of the $|\ell|=2$ manifold which will be of particular interest in this work, the fine structure is presented in Fig. \ref{IntroFig} (c). The highest ($\psi_{1}$) and lowest ($\psi_{4}$) energy levels correspond to linear combinations of $\pm 4\pi$ phase vortices, each associated to orthogonal circular polarizations ($\sigma_{\mp}$). Therefore, these states do not carry a net orbital angular momentum and are linearly polarized, either azimuthally ($\psi_{1}$) or radially ($\psi_{4}$). The middle states ($\psi_{2,3}$) do carry a net angular momentum ($\ell=\pm 2$) and exhibit opposite circular polarizations. It is thus possible to favour gain in either of these chiral modes by spin-polarizing the gain medium with a circularly polarized off-resonant pump\cite{Zambon2018}. In this work, we show the emergence of a bistable regime involving states $\psi_{1}$ and $\psi_{2}$ of this fine structure.
		
		\begin{figure}
			\includegraphics[trim=0cm 0cm 0cm 0cm, width=85mm]{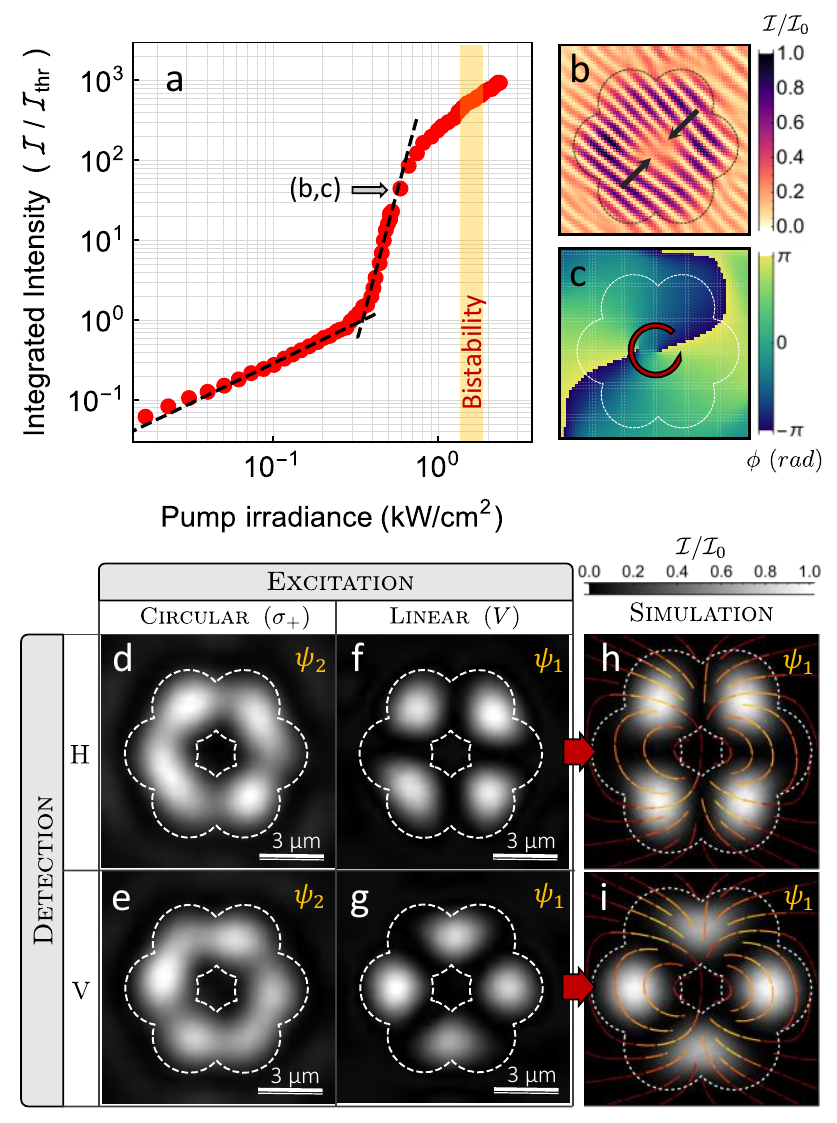}
			\caption{(a) Integrated emission power as a function of the pump density power. The yellow area indicates the bistability region. (b)-(c) Self-interference pattern (b) and extracted phase map (c) of the beam under a $\sigma_{+}$ polarized pump. (d)-(g) Real space images of the beam for a $\sigma_{+}$ (d,e) and horizontal (f,g) polarization pump. The detection is filtered in linear polarization, either horizontal (d,f) or vertical (e,g). Panels (h) and (i) show calculated beam profile of mode $\psi_{1}$, for horizontal (h) and vertical (i) polarization filterings (arrows indicate the average orientation of the electric field).}
			\label{laser}
		\end{figure}
		
		\begin{figure*}
			\includegraphics[trim=0cm 0cm 0cm 0cm, width=0.9\textwidth]{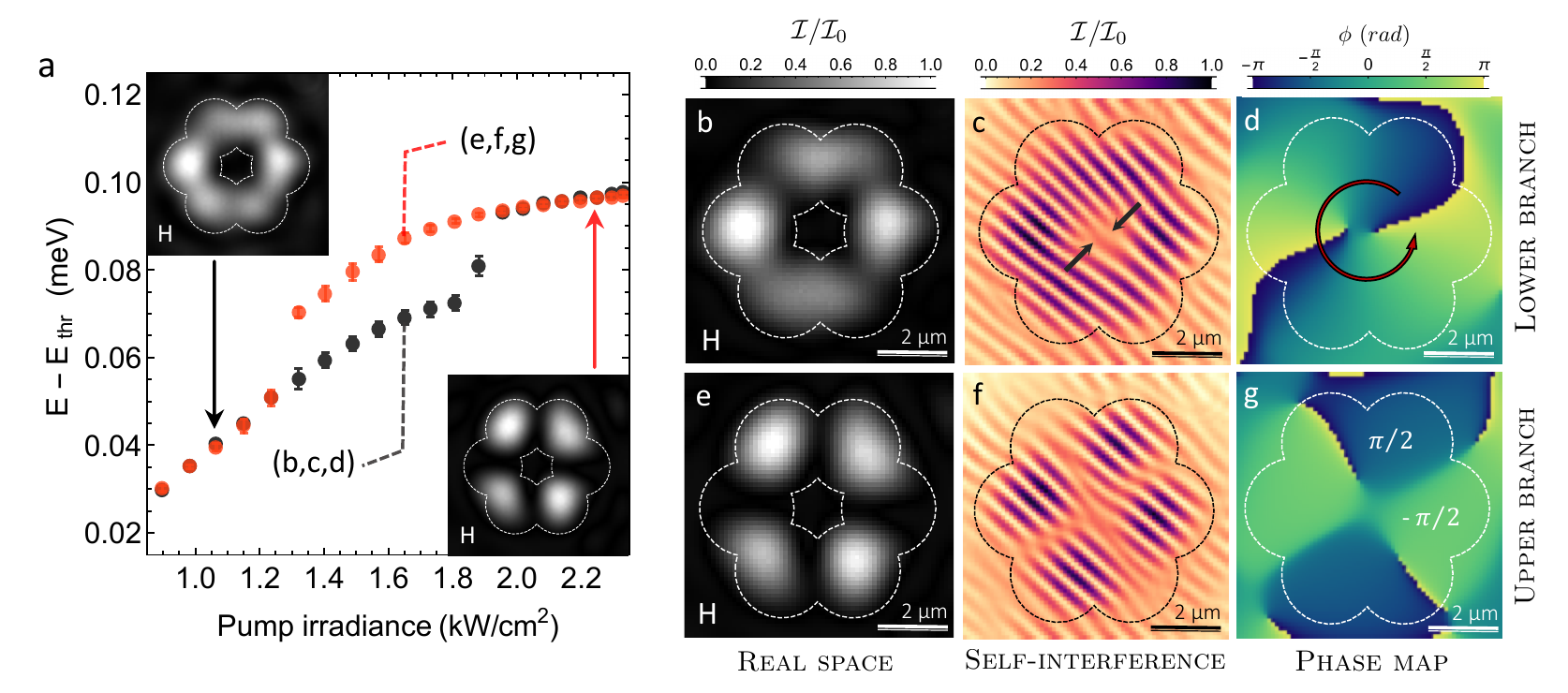}
			\caption{(a) Emission energy as a function of the excitation power, when the power is ramped up (black dots) and down (red dots). The insets show real space images of the beam below and above the bistability. (b)-(g) Real space images (b,e), self-interferometry patterns (c,f) and corresponding phase maps (d,g) measured in the lower (b,c,d) and upper (e,f,g) branches of the bistability, at the same pump power. All images are taken with a horizontal polarization filtering.}
			\label{BistabilityData}
		\end{figure*}
		
		\section{Experimental results} 
		The device investigated presents a geometry such that the gain/loss ratio is maximal for the $|\ell|=2$ manifold (see Ref. \cite{Zambon2018} for details on this lasing scheme). All measurements were done at $\mathrm{T=4~K}$ (details of the experimental setup are furnished in the Suppl. Mat.). The evolution of the emission intensity as a function of the pumping power (Fig. \ref{laser}a) shows a lasing threshold around $\mathrm{P_{th}\sim0.35~kW/cm^{2}}$ and a saturation regime around $\mathrm{0.75~kW/cm^{2}}$. Under a $\sigma_{+}$-polarized pump, lasing occurs in mode $\psi_{2}$ which carries an OAM of $\ell=+2$. This is evidenced by doing a self-interferometry measurement of the beam (Fig. \ref{laser}b) which reveals a double pitchfork in the fringe pattern; as well, the extracted phase map exhibits a $4\pi$ phase vortex (Fig. \ref{laser}c).
		
		Fig. \ref{laser} d-g show real space images of the lasing beam for a circularly ($\sigma_{+}$) and linearly ($V$) polarized pump; the emission is then filtered in linear polarization (either $H$ or $V$). As expected for the $\sigma_{+}$-polarized pump (Panels d and f), the doughnut-like profile of the beam is not significantly affected when changing the detection filtering from one polarization to the other. When pumping the device with a linearly polarized emission, the gain medium is no longer spin-polarized and lasing occurs in the azimuthally polarized mode $\psi_{1}$ (as reported in Ref. \cite{Sala2015}). This mode is identified by spatial images of the beam that reveal four lobes whose position depends on the orientation of the polarization filter (see Panels e and g for experimental data, and Panels h and i for corresponding finite-element simulations).
		
		Upon increasing the incident power of a $\sigma_{+}$-polarized pump far above the lasing threshold, competition between these two modes ($\psi_{1,2}$) leads to the emergence of a bistable regime. An hysteresis cycle is clearly seen in Fig. \ref{BistabilityData} (a) where we present the emission energy as a function of the pump power (the power range of the plot corresponds to the yellow area in Fig. \ref{laser}a). The black (red) dots are measured when the power is ramped up (down). When ramping up the excitation power, the emission energy exhibits an abrupt jump ($\Delta E_{12}\mathrm{\sim20~\mu eV}$) around $\mathrm{P=5.5~P_{th}~(1.85~kW/cm^{2})}$. This jump is accompanied by a drastic change in the spatial profile of the beam: under an horizontal polarization filtering similar to the one used in Fig. \ref{IntroFig} (d) and (e), the profile switches from an homogeneous doughnut shape (upper-left inset) to a four-lobe profile (lower-right inset). Both this shift of energy and change of the spatial profile indicate a mode switch toward the highest energy mode $\psi_{1}$.
		
		Upon decreasing the excitation power (red dots in Fig. \ref{BistabilityData}a), we observe an abrupt lowering of the emission energy around $\mathrm{P=4~P_{th}~(1.3~kW/cm^{2})}$ back to its initial value (i.e. that in the upward scan). This jump is also accompanied by an abrupt change in the spatial pattern, back to its homogeneous shape. We thus evidence an hysteretic cycle that indicates a region of bistability involving two states presenting distinct OAM, i.e. $\ell=0$ ($\psi_{1}$) and $\ell=+2$ ($\psi_{2}$), and distinct polarization textures, i.e. circular ($\psi_{2}$) and azimuthal ($\psi_{1}$) polarizations.
		
		To evidence more clearly this bistable regime, we present images of the beam at an intermediate pump power of $\mathrm{P=4.7~P_{th}~(1.65~kW/cm^{2})}$. Fig. \ref{BistabilityData} presents spatial profiles of the beam (b, e), interferograms (c,f) and corresponding phase maps (d,g), measured respectively in the lower (b-d) and upper (e-g) branches of the hysteresis cycle. Real space images (with horizontal polarization filtering) in the upper and lower branches show respectively the four-lobe and homogeneous patterns characteristic of $\psi_{1}$ and $\psi_{2}$. This identification is further confirmed by interferometry. Fringe patterns measured in the lower branch show two pitchforks as expected for a mode carrying an OAM of $\ell=+2$, and the extracted phase map presents a $4\pi$ phase vortex. When measured in the upper branch, the phase map presents four abrupt jumps between $+\pi/2$ and $-\pi/2$. Such a phase profile describes well the standing wave that characterizes $\psi_{1}$ as a result of the linear combination of counter-propagating components $\ell=+2$ and $\ell=-2$: the phase jumps correspond to the nodes of this standing wave. Therefore, we clearly evidence abrupt switching between modes $\psi_{1}$ and $\psi_{2}$.
		
		\section{Analysis of the bistable regime} 
		In order to describe phenomenologically the emergence of this bistable regime, we use a dynamical model involving rate equations for the time evolution of the two photonic modes ($\psi_{1,2}$) and two reservoir populations ($N_{\uparrow}$ and $N_{\downarrow}$), accounting respectively for spin-up and spin-dwon carriers.
		
		The system is described by the differential equations:
		
		\begin{align}	
			\frac{\mathrm{d}I_{1}}{\mathrm{d}t}=&0.5g_{1}(N_{\uparrow}+N_{\downarrow})I_{1}-\frac{I_{1}}{\tau_{c}}+0.5\frac{N_{\uparrow}+N_{\downarrow}}{\tau_{r}}\nonumber\\
			\frac{\mathrm{d}I_{2}}{\mathrm{d}t}=&g_{2}N_{\uparrow}I_{2}-\frac{I_{2}}{\tau_{c}}+\frac{N_{\uparrow}}{\tau_{r}}\nonumber\\
			\frac{\mathrm{d}N_{\uparrow}}{\mathrm{d}t}=&P(1+\eta)-(0.5g_{1}I_{1}+g_{2}I_{2}+\frac{\beta}{\tau_{r}})N_{\uparrow}\nonumber\\
			\frac{\mathrm{d}N_{\downarrow}}{\mathrm{d}t}=&P(1-\eta)-(0.5g_{1}I_{1}+\frac{\beta}{\tau_{r}})N_{\downarrow}
		\end{align}
		
		\begin{figure}
			\includegraphics[trim=0cm 0cm 0cm 0cm, width=85mm]{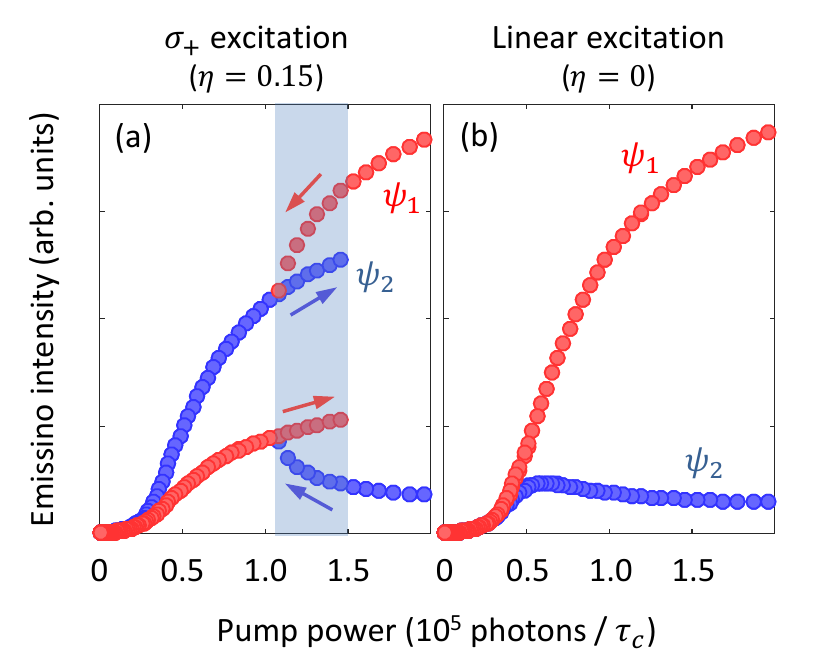}
			\caption{Calculated intensity of the photonic modes $\psi_{1}$ and $\psi_{2}$ under a circularly (a) and linearly (b) polarized excitation. The blue area in Panel (a) corresponds to the bistability region, and arrows indicated the scanning direction of the pump power. The coefficients used in both cases are the following: $g_{0}\mathrm{=11\times10^{-5}~ps^{-1}}$, $\beta\mathrm{=100}$, $\varepsilon_s^{(1)}=\varepsilon_s^{(2)}\mathrm{=5\times10^{-5}}$, $\varepsilon_c^{(1)}\mathrm{=5.43\times10^{-5}}$, and $\varepsilon_c^{(2)}\mathrm{=6\times10^{-5}~ps^{-1}}$.}
			\label{model}
		\end{figure}
		
		Here, $I_{1,2}$ is the photon number in modes $\psi_{1,2}$; $g_{1,2}$ are the gain coefficient of each mode; $\mathrm{\tau_{c}=20~ps}$ is the photon lifetime; $\mathrm{\tau_{r}=100~ps}$ is the carrier lifetime; $\beta$ is the spontaneous emission factor; $P$ is the pump power and $\eta$ is the degree of polarization of the gain medium extracted from the measured degree of polarization of the emission below the lasing threshold.
		
		Bistable regimes have been extensively explored in bimodal lasers, and is attributed to nonlinear contributions to the gain\cite{Tang1987, Kawaguchi1992, Kawaguchi1997}. To account for such effects, we express the gain coefficients as $g_{1,2}=g_{0}(1-\varepsilon_s^{(1,2)}I_{1,2}-\varepsilon_c^{(1,2)}I_{2,1})$, where $g_{0}$ is the unsaturated gain coefficient that we take identical for $\psi_{1,2}$, and $\varepsilon_s^{(1,2)}$ and $\varepsilon_c^{(1,2)}$ are the self- and cross-saturation coefficients of $\psi_{1,2}$. 
		
		For two-mode lasers coupled to a single reservoir, the general requirement for bistability is $\varepsilon_s^{(1)}\varepsilon_s^{(2)}<\varepsilon_c^{(1)}\varepsilon_c^{(2)}$\cite{Tang1987, Kawaguchi1992, Kawaguchi1997}. Here, the situation is slightly more complex as the two modes couple to two distinct reservoirs; moreover, due to their different polarization, they couple differently to each reservoir: $\psi_{1}$ (linearly polarized) couples identically to $N_{\uparrow}$ and $N_{\downarrow}$, whereas $\psi_{2}$ (circularly polarized) couples only to $N_{\uparrow}$. In order to account for the effect of this asymmetric coupling on the nonlinear dynamics of the system we impose a second condition: $\varepsilon_c^{(1)}<\varepsilon_c^{(2)}$.
		
		Fig. \ref{model} (a) shows the adiabatic evolution of the computed intensity mode $\psi_{1}$ (red) and $\psi_{2}$ (blue); we clearly see the emergence of a bistable regime indicated by a blue rectangle. The coefficients (presented in the caption of Fig. \ref{BistabilityData}) were defined in order to obtain a lasing threshold and bistability region at similar powers as experimentally. When changing the degree of polarization to $\eta=0$ (Fig. \ref{model}b), thus simulating a linearly polarized pump, we do not observe any bistability, and the emission is now dominated by $\psi_{1}$ as measured experimentally.
		
		\section{Conclusion and perspectives} 
		In this work, we showed how nonlinear effects in chiral microlasers can lead to a bistable regime involving modes with distinct OAM and polarization patterns. We further showed how dynamical rate equations can capture the essence of this process stemming from the confluence of co- and cross-saturation contributions to the gain. Such a regime is very interesting for implementing optical switches based on the OAM of light, as well as for exploring dynamical processes between phase and polarization vortices exhibiting distinct topological charges.
		
		Finally, it is important to point out that such a bistability is not restricted to the specific values of OAM inspected in this work. Fabricating microlasers with $n$ pillars (with $n$ even and $>4$) could allow implementing similar fine structures as in Fig. \ref{IntroFig} (c) with $\ell=n/2-1$\cite{Zambon2018}. This would lead to bistabilities involving modes with arbitrarily large values of OAM.\\
		
		\noindent\textbf{Funding.} This work was supported by the ERC grant Honeypol, the EU-FET grant PhoQus (PHOQUS-H2020-FETFLAG), the QUANTERA project Interpol (ANR-QUAN-0003-05), the French ANR project Quantum Fluids of Light (ANR-16-CE30-0021) and the Labex CEMPI (ANR-11-LABX-0007) and NanoSaclay (ICQOQS, Grant No. ANR-10-LABX-0035), the French RENATECH network, the CPER Photonics for Society P4S, and the M\'etropole Europ\'eenne de Lille via the project TFlight. P. S.-J. acknowledges financial support from the Marie Sklodowska-Curie individual fellowship ToPol and the National Science and Engineering Research Council of Canada (NSERC). 
		
		The authors acknowledge insightful discussions with O. Bleu, G. Malpuech and D. Solnyshkov.
		
%

\end{document}